\documentclass[12pt]{iopart}
\usepackage{setstack,iopams}
\usepackage{psfig,cite}

\newtheorem{thm}{\bf Theorem} %
\newtheorem{prop}[thm]{\bf Proposition} %

\newcommand{\reals}{\mathbb{R}}

\newcommand{\calP}{\mathcal{P}}
\newcommand{\calM}{\mathcal{M}}
\newcommand{\calH}{\mathcal{H}}

\newcommand{\calR}{\mathcal{R}}
\newcommand{\hcalH}{\hat{\mathcal{H}}}

\newcommand{\hU}{\hat{U}}
\newcommand{\hpsi}{\hat{\psi}}

\newcommand{\htau}{\hat{\tau}}

\newcommand{\ho}{_{\scriptscriptstyle  \mathrm{ho}}}
\newcommand{\hop}[1]{_{{\scriptscriptstyle  \mathrm{ho}},#1}}
\newcommand{\mo}{_{\scriptscriptstyle  \mathrm{mo}}}
\newcommand{\mop}[1]{_{{\scriptscriptstyle  \mathrm{mo}},#1}}
\newcommand{\ptl}{_{\scriptscriptstyle  \mathrm{pt}}}
\newcommand{\ptlp}[1]{_{{\scriptscriptstyle  \mathrm{pt}},#1}}

\newcommand{\tfrac}[2]{{\textstyle\frac{#1}{#2}}}

\newcommand{\up}[1]{^{(#1)}}

\newcommand{\sech}{\mbox{sech}}
\newcommand{\lspan}{\mbox{span}}
\newcommand{\supth}{^{\mathrm{th}}}
\newcommand{\supst}{^{\mathrm{st}}}

\newcommand{\lp}{\left(}
\newcommand{\rp}{\right)}
\begin{document}

\title[Darboux transformation and algebraic deformations]{The Darboux transformation
  and algebraic deformations of shape-invariant potentials.}
\author{D G\'omez-Ullate\dag, N Kamran \ddag$\,\;$ and R
Milson\S}
\address{\dag\ Centre de Recherches Math\'ematiques, Universit\'e de Montr\'eal, (QC)  H3C 3J7
Canada. }
\address{\ddag\
Department of Mathematics and Statistics, McGill University,
Montr\'eal (QC) H3A 2K6 Canada.}
\address{\S\ Department of Mathematics and Statistics, Dalhousie University, Halifax (NS) B3H 3J5 Canada.}
\eads{\mailto{ullate@crm.umontreal.ca},
\mailto{nkamran@math.mcgill.ca}, \mailto{milson@mathstat.dal.ca}}

\begin{abstract}
  We investigate the backward Darboux transformations (addition of a
  lowest bound state) of shape-invariant potentials on the line, and
  classify the subclass of algebraic deformations,  those for
  which the potential and the bound states are simple elementary functions.
  A countable family, $m=0,1,2,\ldots$, of deformations exists for
  each family of shape-invariant potentials.  We prove that the
  $m\supth$ deformation is exactly solvable by polynomials, meaning
  that it leaves invariant an infinite flag of polynomial modules
  $\calP\up{m}_m\subset\calP\up{m}_{m+1}\subset\ldots$, where
  $\calP\up{m}_n$ is a codimension $m$ subspace of $\langle
  1,z,\ldots,z^n \rangle$.  In particular, we prove that the first
  ($m=1$) algebraic deformation of the shape-invariant class is
  precisely the class of operators preserving the infinite flag of
  exceptional monomial modules $\calP\up{1}_n = \langle
  1,z^2,\ldots,z^n\rangle$.
  By construction, these algebraically deformed Hamiltonians do not
  have an $\mathfrak{sl}(2)$ hidden symmetry algebra structure.
\end{abstract}
 \pacs{03.65.Fd, 03.65.Ge}

\newcommand{\anote}[1]{[{\small\em #1}]}

\newcommand{\czeta}{c\kern 1pt\zeta}

\maketitle
\section{Introduction.}
The Darboux transformation\cite[p.~210]{darboux}\cite{jacobi}, a
method based on the factorization of second-order operators, is an
important technique for the exact solution of the one-dimensional
Schr\"odinger equation\cite{schrodinger,infeld-hull}.  The
transformation is also a key concept in supersymmetric quantum
mechanics\cite{cooper} and the theory of integrable systems
\cite{deift,gesztesy,calogero}.  From the point of view of
spectral theory, a non-singular Darboux transformation can be
characterized by the following three
possibilities\cite{sukumar}.  First, a potential with a lowest
bound state admits a unique forward Darboux transformation, which
deletes the ground state. Second, a potential admits a
2-parameter family (the energy and a shape parameter) of backward
Darboux transformations, each of which adds a lowest eigenvalue.
Third, there is also a 1-parameter family of isospectral
transformations, corresponding to two critical values of the shape
parameter, which neither add nor delete bound
states\cite{sparenberg}.


The Darboux transformation relates to exact solutions in several
ways\cite{bagrov1}.  Generally, the forward transformation
deforms a given exactly solvable potential to a new, solvable
potential form. However, if a parameterized family of exactly
solvable potentials is shape-invariant\cite{gendenshtein}, i.e.,
closed with respect to the forward Darboux transformation, then
the forward transformation furnishes an explicit description of
the spectrum and eigenfunctions, rather than a new
exactly-solvable form.  Therefore, to obtain a solvable
deformation of a shape-invariant potential it is necessary to
employ a backward Darboux transformation. This was first done for
specific potentials such as the harmonic
oscillator\cite{mielnik}, while the general theory was developed
in \cite{deift,sukumar}.

However, the general form of the deformed potential features integrals
of eigenfunctions of the original Hamiltonian --- in contrast to the
original potential, which is an elementary function, with bound states
also described by elementary functions.  It has been noted\cite{levai}
that only certain discrete values of the energy and shape parameter
correspond to an \emph{ algebraic deformation}, one where the
potential and the bound states remain elementary functions.  Such
forms are attractive from the modeling standpoint, and are also
important theoretically, since exact results can be obtained even in
critical conditions, where numerical techniques break down.

In the present article we explain the discrete nature of algebraic
deformations by characterizing such deformations in terms of
polynomial modules left invariant by a second-order differential
operator.  The invariant module approach is an alternative, inherently
algebraic, approach to exact solvability --- developed originally to
treat quasi-exactly solvable Hamiltonians \cite{turbiner1,ko,gko}. In
this approach, one considers a Hamiltonian that, in a suitable gauge,
preserves an infinite flag of polynomial modules
\begin{equation}\label{pflag}
\calP_0\subset\calP_1 \subset\calP_2\subset\ldots\subset
\calP_n\subset\ldots,\qquad \calP_n = \langle 1,z,\ldots,z^n\rangle.
\end{equation}
Such a Hamiltonian has an upper-triangular action, and it is, therefore,
algebraically diagonalizable.
On the line, there are exactly three potential forms whose
Hamiltonian, in a suitable gauge, preserves \eref{pflag}.  These are
the classical shape-invariant potential families: the harmonic
oscillator, the Morse\cite{morse}, and the hyperbolic
P\"oschl-Teller\cite{ptell}  potentials.
This invariant module approach provides an alternative explanation for
the exact-solvability of these shape-invariant potentials.

The details of this approach can be found in \cite{turbiner1,gko}, and some
generalizations in \cite{milson}. The method of invariant polynomial
flags has also been applied in quantum many-body problems \cite{ggr}.
The question of determining which second-order operators preserve a
finite-dimensional polynomial module has been previously considered in
a number of papers, including \cite{turbiner2,finkelkamran}.

We consider the algebraic deformations of the three shape-invariant
potential families \cite{bagrov1,levai,baye}, and show that, in each
case, the $m\supth$ algebraic deformation produces a Hamiltonian that,
modulo a gauge transformation and a change of variable, preserves an
infinite flag of deformed polynomial modules
\begin{equation}
  \label{eq:gapmodflag}
  \calP\up{m}_{m}\subset \calP\up{m}_{m+1}\subset\calP\up{m}_{m+2}\subset\ldots\subset
  \calP\up{m}_{n}\subset\ldots,
\end{equation}
where each $\calP\up{m}_n$ is a certain codimension $m$ subspace of
$\calP_n$, i.e. the span of $n-m+1$ polynomials of degree $n$.

We study the first deformation ($m=1$) in some detail, and show that
$\calP\up{1}_n$ is an \emph{exceptional monomial
module}\cite{turbiner2}, an invariant vector space spanned by
\[1,z^2,z^3,\dots,z^n \qquad\mbox{(the first power is omitted)}.\]
We also show that higher deformations produce more complicated
polynomial modules, but we do not analyze these modules here.

We will discuss monomial modules more thoroughly in a forthcoming
publication \cite{gkm}. At this point, we would like to mention that
exceptional monomial modules also arise in the context of N-fold
supersymmetry, \cite{gt}. Our emphasis is somewhat different, since we
are primarily concerned with the interplay between the backward
Darboux transformation and the class of the operators preserving an
infinite flag of polynomial modules.

We also note that invariance of \eref{pflag} is generally
achieved by expressing the gauge hamiltonian as a quadratic
combination of those generators of $\mathfrak{sl}(2)$, realized
as first order differential operators, which leave invariant the
infinite flag of polynomial subspaces $\calP_n$.  These operators
are called {\em Lie-algebraic} and the Hamiltonian is said to
have a hidden $\mathfrak{sl}(2)$ symmetry algebra. Lie-algebraic
potentials in one dimension have been classified in \cite{gko}.
However, not all exactly-solvable potentials are generated by a
hidden symmetry, the Coulomb potential being a notable
counterexample\cite{shifman}.  Since the potentials studied here
preserve \eref{eq:gapmodflag} rather than \eref{pflag}, they lack
an $\mathfrak{sl}(2)$ hidden symmetry algebra structure, and thus
furnish a further indication that the exactly solvable class of
potentials is larger than the Lie algebraic one
\cite{turbiner1,ko,gko}.

This paper is structured as follows. In the next Section we describe
the forward and backward Darboux transform.  In Section
\ref{deformations} we discuss exactly solvable operators and algebraic
deformations. We also exhibit the invariant flags corresponding to the
algebraic deformations of the three shape-invariant potential
families. In Section $4$ we consider exactly solvable operators that
preserve the exceptional monomial module, and demonstrate that these
are precisely the first-fold deformations of the shape-invariant
Hamiltonians discussed in Section \ref{deformations}.


\section{Factorization and the Darboux transformation}\label{darboux}
Let $U(x),\; x\in(-\infty,\infty)$ be a continuous, real-valued function,
and let
\[\tau = -\partial_{xx}+U,  \]
be the corresponding formally self-adjoint differential operator.
We fix a formal  eigenfunction
\[\tau[\phi]=\lambda_0\, \phi,\]
and note that every such $\phi$ corresponds to a differential
factorization
\[\tau -\lambda_0 = \alpha^* \alpha,\]
where
\begin{equation}
  \label{eq:factor-def}
  \alpha = \partial_x+\sigma_x,\quad \alpha^*=-\partial_x+\sigma_x,\quad
  \sigma=-\ln \phi.
\end{equation}
For this reason we shall refer to $\phi$ as the \emph{factorization function},
and to $\lambda_0$ as the \emph{factorization energy}.

Commutation of the factors defines a partner differential operator
\[\htau =\alpha\alpha^*=-\partial_{xx}+ \hU,\]
where
\[\hU= U+ 2 \sigma_{xx}.\]
The operators obey the following intertwining
relation:
\begin{equation}
  \label{eq:intertwine}
  \alpha\tau=\htau\alpha.
\end{equation}
As a consequence,
the first-order operator $\alpha$ relates the
eigenfunctions of the two operators:
given
\[\tau[\psi]=\lambda \psi,\]
we also have
\begin{equation}
  \label{eq:dxform}
  \htau[\hpsi] = \lambda \hpsi,\quad \hpsi=\alpha[\psi].
\end{equation}

To give a rigorous treatment of the Darboux
transformation\cite{gesztesy,schminke}, we assume that $U(x)$ is
bounded from below, and let $\calH$ be the unique self-adjoint,
semi-bounded operator corresponding to $\tau$.  The partner
potential $\hU(x)$ is continuous if and only if $\phi$ is
non-vanishing.  In this case, the spectrum of $\hU(x)$ is bounded
from below, and we let $\hcalH$ denote the corresponding
self-adjoint, semi-bounded operator. Letting $A$ denote the
closed operator corresponding to $\alpha$, we have that
$\alpha^*$ corresponds to the adjoint $A^*$.  We therefore obtain
the following non-formal factorizations:
\[ \calH -\lambda = A^* A,\quad \hcalH-\lambda = A A^*,\]
where the compositions are appropriately restricted.

In particular, $A$ maps $D(\calH)$, the domain of $\calH$, to
$D(\hcalH)$. The spectral properties of this transformation are
governed by one of the following 3 possibilities\cite{deift}.
\begin{enumerate}
\item {\bf Forward transformation:} $\phi$ is square integrable, and
  $A$ defines an isomorphism between $\ker A=\phi^\perp$ and $D(\hcalH)$.
  Thus, a forward transformation exists if and only if $\calH$
  possesses a discrete spectrum, in which case it is unique.  The
  transformed spectrum differs from  the spectrum of $\calH$ by the removal of
  $\lambda$, the lowest eigenvalue.  The first-order operator $\alpha$
  transforms the $n\supth$ bound state of $\calH$ to the $(n-1)\supst$
  bound state of $\hcalH$.
\item {\bf Backward transformation:} $\phi^{-1}$ is square integrable,
  and $A$ defines an  embedding of $D(\calH)$ into
  $D(\hcalH)$.  The range of the embedding is precisely
  $(\phi^{-1})^\perp$.  The spectrum of $\hcalH$ differs from that of
  $\calH$ by the addition of  a lowest eigenvalue, namely $\lambda$, with
  the ground state given by $\phi^{-1}$.
  The operator $\alpha$
  transforms the $n\supth$ bound state of $\calH$ to the $(n+1)\supst$
  bound state of $\hcalH$.

  A 1-parameter family of backward transformations exist for every
  $\lambda$ strictly smaller than the spectrum of $\calH$.  To
  describe the possibilities, let $\phi_+$ and $\phi_-$ be the unique
  (up to a multiple) positive solutions of
  \begin{equation}
    \label{eq:phi0def}
    \tau[\phi_\pm]=\lambda\phi_\pm,
  \end{equation}
  with the property that $\phi_\pm$ is square
  integrable near, respectively,  $\pm\infty$.  The desired $\phi$ is
  of the form
  \[\phi = s\phi_++t\,\phi_-,\quad s,t>0.\]
\item {\bf Isospectral transformation:} neither $\phi$ nor $\phi^{-1}$
  are square integrable.  In this case $A$ acts as an isomorphism between
  $D(\calH)$ and  $D(\hcalH)$.    The operator $\alpha$
  transforms the $n\supth$ bound state of $\calH$ to the $n\supth$
  bound state of $\hcalH$. Two isospectral Darboux
  transformations exist for every $\lambda$ strictly smaller than the
  spectrum of $\calH$: one corresponding to $\phi=\phi_+$, and the
  other to $\phi=\phi_-$.
\end{enumerate}

\section{Algebraic deformations of shape-invariant
  potentials.}\label{deformations}
\subsection{Exact solvability}
We will call a Schr\"odinger operator
\[\calH=-\partial_{xx} + U,\]
\emph{exactly solvable by polynomials} if $\calH$ is equivalent,
by a change of variable and a gauge transformation, to a
second-order operator $T$ that preserves an infinite flag of
finite-dimensional polynomial modules
\begin{equation}
  \label{eq:mflag}
  \calM_1\subset \calM_2\subset\calM_3\subset\ldots,\quad
  \dim \calM_n = n.
\end{equation}
The exact solvability comes about because $T$ is
upper-triangular relative to a basis adapted to the above flag, and
hence possesses an infinite list of eigenpolynomials.
In this paper we will make the hypothesis that such
an operator is  of the form
\[ T = P(z)\partial_{zz} + Q(z) \partial_{zz} + R(z),\]
where
\[ P(z) = p_2 z^2 + p_1 z + p_0,\quad p_0,p_1,p_2\in\reals,\]
is a polynomial of degree $2$ or less, and where $Q(z), R(z)$ are
rational functions.

We transform the second-order eigenvalue equation
\[ T[f]= \lambda f\]
into a Schr\"odinger equation
\[\calH[\phi]= \lambda \phi\]
by a change of variables
\begin{equation}
  \label{eq:varchange}
  x = \int^{z} (-P)^{-\frac{1}{2}} ,
\end{equation}
and a  gauge transformation
\begin{equation}
  \label{eq:gaugexform}
  \phi = e^{p} f\Big|_{z=z(x)}
\end{equation}
where
\begin{eqnarray}
  \label{eq:gaugefac}
    p &=& \int^z
  \tfrac{1}{2} P^{-1}(Q-\tfrac{1}{2}P'),\\
  \label{eq:potform}
  U &=& \tfrac{1}{4}P''-\tfrac12 Q' -\tfrac14P^{-1}\lp Q-\tfrac12
  P')(Q-\tfrac32 P'\rp+R\Big|_{z=z(x)}
\end{eqnarray}
Let $\calR$ denote the range of the function $z(x)$, and let us assume
that neither $Q(z)$ nor $R(z)$ have singularities for $z\in \calR$.
We assume that $P(z)$ is not everywhere positive.  If it were, then we
change $P\to-P$.  With these assumptions we prove the following
\begin{prop}
\label{prop:zrange}
Exactly one of the following possibilities holds, according to the
number and multiplicity of the real roots of $P(z)$:
\begin{enumerate}
\item The are no real roots. Then,
  $\calR=(-\infty,\infty)$, and $U(x)$ is non-singular.
\item There is a double root, $\rho$. Then,
  $\calR=(\rho,\infty)$ or $(-\infty,\rho)$, and $U(x)$ is
  non-singular.
\item There is a unique, simple root, $\rho$. Then, $\calR
  =(-\infty,\rho]$, or $[\rho,\infty)$.  Both $z(x)$ and $U(x)$ are
  even functions. The potential is non-singular if and only if
  \begin{equation}
    \label{eq:zrange}
    Q(\rho)= \tfrac12 P'(\rho),\quad\mbox{or}\quad Q(\rho)=\tfrac32 P'(\rho).
  \end{equation}
\item There are two distinct roots $\rho_1<\rho_2$, and $\calR
  =(-\infty,\rho_1]$, or $[\rho_2,\infty)$. Then, both $z(x)$ and the
  potential are even functions. The potential is non-singular if and
  only if \eref{eq:zrange} holds with, $\rho=\rho_1$, or with
  $\rho=\rho_2$, respectively.
\item There are two distinct roots $\rho_1<\rho_2$, and
  $\calR=[\rho_1,\rho_2]$.
  Then, both $z(x)$ and the potential are periodic functions, and
  $U(x)$ has a singularity.
\end{enumerate}
\end{prop}
\emph{Proof.} Condition \eref{eq:zrange} follows from
\eref{eq:potform}.  In order for the potential to be nonsingular, the
function $( Q-\tfrac12 P')(Q-\tfrac32 P')$ must vanish at those roots
of $P(z)$ that lie in $\calR$.  The singularity occurs in case {\it
  (v)}, because both $Q(z)$ and $P'(z)$ are first-degree polynomials,
and hence \eref{eq:zrange} cannot hold for both $\rho=\rho_1$ and
$\rho=\rho_2$.


\subsection{Shape-invariant potentials on the line.}
Let us begin by showing, that the class of exactly solvable
Hamiltonians that preserve the infinite flag \eref{pflag} consists of
shape-invariant potentials on the line.
Up to a constant, the most general form of a second-order operator
that preserves \eref{pflag} is
\begin{equation}
  \label{eq:esform}
  T=P(z)\partial_{zz} + Q(z) \partial_z,
\end{equation}
where $P(z)$ and $Q(z)$ are arbitrary second and first degree
polynomials, respectively.  Since $\calP_n$ is invariant with respect
to affine transformations of the $z$ variable, we can use such
transformations to modify $P(z)$ and $Q(z)$ without loss of
generality.  Additionally, by rescaling the $x$ variable, and shifting
the spectrum, we reduce our analysis to one of the following four
canonical forms:
\begin{table}[htbp]
  \begin{center}
    \begin{tabular}{c|cccc}
      & $P(z)$ & $Q(z)$ & $z(x)$ & $U(x)$\\
      \hline
      Ia & $ -1$ & $2z$& $x$   & $x^2$\\
      Ib& $-4z$ &$4z-2$& $x^2$ & $x^2$,\\
      II & $-z^2$ & $-(2A+3)z+1$ & $e^{x}$ & $\tfrac14 e^{-2x}
      -(A+\tfrac12) e^{-x}$,\\
      III & $z(1-z)$ & $(A-\tfrac32)z+1-A$& $\cosh^2(\tfrac{x}{2})$ &
      $\tfrac14(\tfrac14-A^2)\sech^2(\tfrac{x}{2})$.
    \end{tabular}
    \caption{Shape-invariant potentials on the line}
    \label{tab:si}
  \end{center}
\end{table}

By Proposition \ref{prop:zrange}, the $Q(z)$ in cases Ia, Ib, and III is
the most general form for which $U(x)$ is non-singular.  In case II,
we use a translation of the $x$ variable to fix the form of
$Q(z)$. Thus, we have
derived the classical shape-invariant potential families: the harmonic
oscillator (Ia) (Ib), the Morse potential (II), and the hyperbolic
P\"oschl-Teller potentials (III). Each of these potential forms is
discussed in more detail below.


Let us now show that these potentials are shape-invariant by
construction.  A parameterized family of potentials is called
shape-invariant if the ground state Darboux transformation acts by
changing the potential parameters, but leaves the form of the
potential invariant.  For the operators in question, the ground state
energy is $\lambda=0$, and hence the corresponding factorization is
given by
\[\calH= \lp -(-P)^{\frac{1}{2}} \partial_z
+\tfrac12(Q-\tfrac12 P') (-P)^{-\frac{1}{2}}\rp \lp  (-P)^{\frac{1}{2}}
\partial_z+\tfrac12(Q-\tfrac12 P') (-P)^{-\frac{1}{2}}\rp \Big|_{z=z(x)},
\]
where $z=z(x)$ is the change of variable defined in
\eref{eq:varchange}.  The commutation of the factors produces a
Schr\"odinger operator $\hcalH$, which corresponds to the algebraic
operator
\[ \hat{T} = P(z)\partial_{zz}+\hat{Q}(z)\partial_z,\]
where
\begin{equation}
  \label{eq:hatq}
  \hat{Q}=P'-Q.
\end{equation}
For this reason, the forward Darboux transformation for these
potentials produces another potential \eref{eq:potform} of the
same form, but with potential parameters modified by
\eref{eq:hatq}.

\subsection{Algebraic deformations.}
Now, let us isolate the values of the energy and shape parameters for
which the backward Darboux transformation of the shape-invariant
potentials yields an algebraic deformation.  We will say that a
Darboux transformation is an \emph{algebraic deformation}, when the
derivative of $\ln \phi$ in \eref{eq:factor-def} is either a rational
function, or a composition of a rational function with an exponential.
As per \eref{eq:gaugexform}\eref{eq:esform}, for the shape-invariant
potentials under discussion, the factorization function is of the form
\[\phi(x)=e^{p(z(x))} f(z(x)),\]
where $p(z)$ is a polynomial,  $z(x)$ is an elementary
function, and $f(z)$ is a hypergeometric function.  We will say
that such an $f(z)$ is of \emph{polynomial type} if $f'(z)/f(z)$ is a
rational function.  We observe that
\[ (\ln \phi(x))' = p'(z(x)) z'(x) + \frac{f'(z(x))}{f(z(x))} \,
z'(x).\]
Thus, to obtain an algebraic deformation we
must demand that $f(z)$ be of polynomial type, with $f(z)$ non-vanishing in
the range $\calR$ of $z(x)$ (see Proposition \ref{prop:zrange}).

The regular solution of the confluent hypergeometric
equation\cite[Sec. 6.1]{bateman},
\begin{equation}
  \label{eq:chg}
 z y''(z) + (c-z) y'(z) - a y(z) =0,
\end{equation}
is given by the confluent hypergeometric function
\[y(z) = \Phi(a,c,z) = {}_1F_1(a,c;z).\]
It can be shown \cite[Sec. 6.9]{bateman} that solutions of
polynomial type exist if and only if either $a$ or $c-a$ is an
integer.  These solutions, expressed in terms of generalized
Laguerre polynomials $L_m^a(z),\; m=0,1,2,\ldots$, are given
below:
\begin{equation}
  \label{eq:chgpolys}
  \begin{array}{rll}
    y_1(z)&=\Phi(a,c;z)\quad &a=-m,\\
    &\propto L_m^{c-1}(z) \\
    y_2(z)&=z^{1-c}\,\Phi(a-c+1,2-c;z)\quad &c-a=1+m,\\
    &\propto z^{1-c} L_m^{1-c}(z)\\
    y_3(z)&=e^z\Phi(c-a,c;-z)\quad &c-a=-m,\\
    &\propto e^z L_m^{c-1}(-z)\\
    y_4(z)&= z^{1-c}e^z\, \Phi(1-a,2-c,-z)\quad &a=m,\\
    &\propto z^{1-c} e^z L_m^{1-c}(-z)
\end{array}
\end{equation}
In general, we have $y_1(z)=y_3(z)$ and $y_2(z)=y_4(z)$
\cite[Sec. 6.4]{bateman}.

The regular solution of the hypergeometric equation\cite[Sec.
6.1]{bateman},
\begin{equation}
  \label{eq:hg}
 z(1-z) f''(z) + (c-(a+b+1)z) f'(z) - ab f(z) =0,
\end{equation}
is given by the Gauss hypergeometric function
\[f(z) = F(a,b,c;z)={}_2F_1(a,b,c;z).\]
Solutions of polynomial type exist if and only if \eref{eq:hg} is
of so-called degenerate type.  This means that the monodromy
group around one of the regular singular points $0,1,\infty$ is
trivial\cite[Sec. 2.2]{bateman}. These solution, expressed in
terms of Jacobi polynomials $P_m^{(\alpha,\beta)}(z)$, are shown
below:
\begin{equation}
  \label{eq:hgpolys}
  \hskip-5em\begin{array}{rll}
    f_1(z)&=F(a,b,1+a+b-c;1-z),\; &a=-m,\\
    & \propto P_m^{(\alpha,\beta)}(2z-1) &\alpha=b-c-m,\;
    \beta=c-1\\
    f_2(z)&=z^{1-c}(1-z)^{c-a-b} \,F(1-a,1-b,1-a-b+c;1-z)
    &a=m+1,\\
    &\propto z^\beta (1-z)^\alpha\,
    P_m^{(\alpha,\beta)}(2z-1)&\alpha=c-b-1-m,\;\beta=1-c\\
    f_3(z)&=(1-z)^{c-a-b} F(c-a,c-b,1-a-b+c;1-z) &c-a=-m,\\
    &\propto (1-z)^\alpha  P_m^{(\alpha,\beta)}(2z-1),&\alpha=-b-m,\;
    \beta=c-1\\
    f_4(z)&=z^{1-c}F(a+1-c,b+1-c,1+a+b-c;1-z) &c-a=m+1,\\
    &\propto z^\beta P_m^{(\alpha,\beta)}(2z-1),&\alpha=b-1-m ,\;
    \beta=1-c
\end{array}
\end{equation}

We will need \eref{eq:chgpolys} to construct algebraic deformations of
the Harmonic oscillator and the Morse potential, and \eref{eq:hgpolys}
to construct the algebraic deformations of the hyperbolic
P\"oschl-Teller potential.
\subsection{The Harmonic oscillator.}
The general solution of
\begin{equation}
  \label{eq:hoformals}
  -\phi''(x) + U\ho(x)\phi(x) = \lambda\phi(x),\qquad U\ho(x) =
x^2,
\end{equation}
is given by
\begin{equation}
  \label{eq:hosols}
  \phi\ho(x;\lambda,A_0,A_1) = \lp A_0\, \Phi\!\lp \textstyle
  \frac{1}{4}-\frac{\lambda}{4},\frac{1}{2};x^2\rp
  +A_1\, x\,\Phi\!
\lp{\textstyle\frac{3}{4}-\frac{\lambda}{4},\frac{3}{2},x^2}\rp\rp
  \e^{-\frac{x^2}{2}},\nonumber
\end{equation}
The $n\supth$ bound state is
\begin{equation}\label{phis}
  \psi\hop{n}(x)\propto e^{-\frac{x^2}{2}} H_n(x) \propto \left \{
    \begin{array}{ll}
      \phi\ho(x;1+2n,1,0), & n
    \mbox{ even,} \\
     \phi\ho(x;1+2n,0,1), & n \mbox{ odd,}
    \end{array}\right.
\end{equation}
where $H_n(z)$ denotes the $n\supth$ Hermite polynomial. For $\lambda<1$, the
nodeless solutions of \eref{eq:hoformals}
are given by
\begin{equation}
  \label{eq:hosc+-}
  \phi\ho(x;\lambda,1,t),\quad |t|\leq 2\,\frac{\Gamma(\frac{3}{4} -
    \frac{\lambda}{4})}{\Gamma(\frac{1}{4} - \frac{\lambda}{4})},
\end{equation}
with the extreme values of $t$  corresponding to $\phi_\pm$. This
follows from the asymptotic properties of $\Phi$ for large $x$
\cite[Sec. 6.13.1]{bateman}.

Applying \eref{eq:chgpolys} with $c=1/2$ and $a=1/4-\lambda_0/4$, c.f.
\eref{eq:hosols}, it follows that algebraic deformations only occur
when the factorization energy $\lambda_0$ is an odd integer.  We rule
out $y_1(z)$ and $y_2(z)$ because we need $\lambda_0<1$.  We rule out
$y_4(z)$, because \eref{eq:hosc+-} shows that the corresponding
eigenfunction always has node.  Thus, we are left with factorization
functions of the form
\[\phi\up{m}\ho(x)=e^{-\frac{x^2}{2}} y_3(x^2)=\phi\ho(x;-1-4m,1,0) \propto
 e^{\frac{x^2}{2}}H_{2m}(ix).\]
In this way
we obtain the following algebraic deformations of the harmonic
oscillator:
\begin{equation}
  \label{eq:modhosc}
  U\up{m}\ho(x) = x^2 - 2\,\partial_{xx} (\log H_{2m}(ix))-2,\quad
  m=0,1,2,\ldots.
\end{equation}

\begin{figure}[htbp]
  \begin{center}
    \noindent\psfig{figure=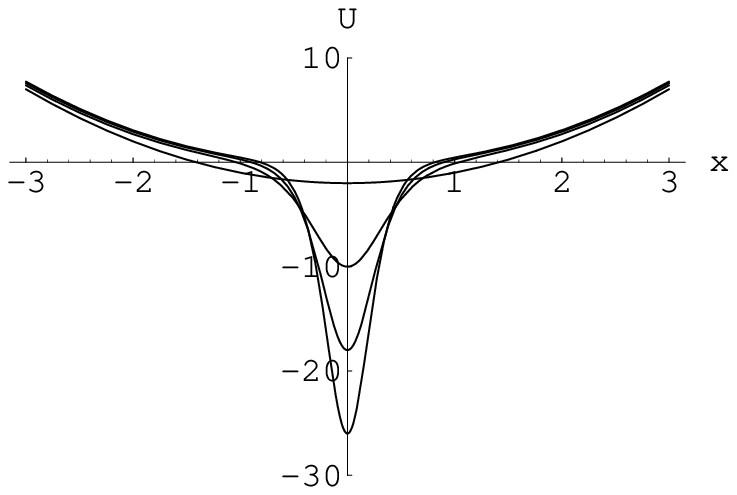,width=4in}
    \caption{Deformations $U\up{m}\ho(x)$ of the harmonic oscillator for $m=0,1,2,3$}
    \label{fig:ho}
  \end{center}
\end{figure}

The harmonic oscillator is shape invariant, and so the zeroth
deformation is again a harmonic oscillator, albeit with a spectral
shift.  The potentials and eigenfunctions of the higher deformations
are described in \cite{dubov,bagrov2}.  The full, two parameter family
of backward transformations is discussed in \cite{sukumar}.

The new spectral minimum is $-1-4m$, and the new ground state is a
 multiple of
\[    \psi\up{m}\hop{0}(x)\propto(\phi\up{m}\ho(x))^{-1}\propto e^{-\frac{x^2}{2}}(H_{2m}(ix))^{-1}.\]
The rest of the spectrum is unchanged. The Darboux transformation
corresponds to the operator
\begin{eqnarray*}
   \alpha\ho\up{m} &=& \partial_x -\partial_x\log\left(
     e^{\frac{x^2}{2}}H_{2m}(ix)\right) \\
   &=&\partial_x -x -\frac{4im H_{2m-1}(ix)}{H_{2m}(ix)}.
 \end{eqnarray*}
Consequently, the higher bound states
are
\[  \psi\up{m}\hop{j}\propto  \alpha\ho\up{m}\,[\psi\hop{j-1}] =
  e^{-\frac{x^2}{2}}(H_{2m}(ix))^{-1}\, p\up{m}_j(x),\quad j=1,2,\ldots
\; ,\]
where
\begin{eqnarray*}
  p\up{m}_j(x) &=&
2(j-1)H_{2m}(ix)H_{j-2}(x) -2x H_{2m}(ix) H_{j-1}(x)\\
  &&- 4im\,  H_{2m-1}(ix)H_{j-1}(x).
\end{eqnarray*}
Thus, the even polynomials
\[q_k(z)=p\up{m}_{2k}(x) ,\quad z=x^2,\quad k=1,\ldots,n\]
together with $q_0=1$, span an invariant $(n+1)$-dimensional
submodule of $\calP_{n+m}$.  The odd polynomials
\[r_k(z) = x^{-1} p\up{m}_{2k+1}(x),\quad k=0,\ldots,n\]
also span an $(n+1)$-dimensional invariant submodule of
$\calP_{n+m}$.  Therefore, algebraic deformations of the harmonic
oscillator are  exactly solvable by polynomials.
\subsection{The Morse potential.}
\label{sub:morse}
The  Morse potential\cite{morse} has the form
\begin{equation}\label{Morse}
U\mo(x) =  -(A+\tfrac{1}{2}) e^{-x} +  \tfrac{1}{4}e^{-2x}.
\end{equation}
The function
\begin{equation}
  \label{eq:morsegs}
  \phi\mo(x;k,C_+,C_-) = \sum_{\pm} C_{\pm} e^{\pm
  kx-\frac{1}{2}e^{-x}} \Phi(\mp
  k-A,1\mp 2k,e^{-x})
\end{equation}
is, generically, the general solution of the corresponding
Schr\"odinger equation
\[ -\phi''(x) + U\mo(x)\phi(x)  = -k^2\phi.\]
In the singular case where $1-2k$ is a non-positive integer, the
general solution can be given as
\begin{eqnarray}
  \label{eq:morsegs1}
 \phi\mo(x;k,C_+,C_-) &=&   C_+\,
 e^{kx-\frac{1}{2}e^{-x}}\Psi(-k-A,1-2k,e^{-x})\\
 &&+\,C_-\,e^{ -kx-\frac{1}{2}e^{-x}}\Phi(k-A,1+2k,e^{-x}), \nonumber
\end{eqnarray}
where $\Psi$ is the irregular solution of the confluent hypergeometric
   equation.

There are no bound states if $A\leq 0$, and $\lceil A \rceil$ bound
states otherwise, with
the $n\supth$ bound state is
\begin{equation}
  \label{eq:morse-bound}
  \hskip -3em \psi\mop{n}(x)\propto\phi\mo(x,A-n,0,1)\propto e^{(n-A)x-\frac{1}{2}e^{-x}}
  L_n^{2(A-n)}(e^{-x}),\quad 0\leq n<A.
\end{equation}
We will focus on deformations of potentials with bound states only.
For $A>0$, the spectral minimum is $-A^2$, and hence we must have
\begin{equation}
  \label{eq:mo-kA}
  A<|k|.
\end{equation}
In the non-singular case we have by~\cite[Sec.~6.13.1]{bateman}
\begin{eqnarray}
\label{eq:psi+mo}
\phi_+&=&\phi\mo(x;k,0,1),\\
\label{eq:psi-mo}
\phi_-&=& \phi\mo\lp
x;k,1,-\frac{\Gamma(1-2k)}{\Gamma(1+2k)}\frac{\Gamma(k-A)}{\Gamma(-k-A)}\rp.
\end{eqnarray}
In the singular cases, when
$2k-1$ or $A+k$ is $0,1,2,\ldots$,
we  have
\begin{equation}
  \label{eq:psi-mos}
  \phi_-=\phi\mo(x;k,1,0).
\end{equation}
The above hold for $k>A$.  For $k<-A$, the order of $C_+, C_-$ is
reversed.

To obtain algebraic deformations we apply \eref{eq:chgpolys} with
$a=-k-A$ and $c=1-2k$, and consider the four possible factorization
functions
$$\phi_i(x) =  e^{kx-\frac12 e^{-x}}y_i(e^{-x}),\quad i=1,2,3,4,$$
in turn.

Let $m$ be a non-negative integer.  For $\phi_1$ we need $k+A=m$,
which by \eref{eq:mo-kA} implies that $A<\tfrac{m}{2}$.
Hence, by \eref{eq:morsegs} \eref{eq:psi-mos},
\[ \phi_1(x) = \phi\mo(x;m-A,1,0)=\phi_-(x),\]
and therefore it generates an isospectral deformation.

For $\phi_2(x)$ we need $1-k+A=m$, which by \eref{eq:mo-kA} implies
$A<\tfrac{m}{2}-\tfrac{1}{2}$. Hence, by \eref{eq:morsegs}
\eref{eq:psi+mo},
\[ \phi_2(x) = \phi\mo(x;1+A-m,0,1) = \phi_+(x),\]
and therefore it also generates an isospectral deformation.

For $\phi_3(x)$ we need $k-1-A=m$, and hence by \eref{eq:morsegs}
\[ \phi_3(x) = \phi\mo(x;m+1+A,1,0).\]
By \eref{eq:psi-mo} this function is nodeless if and only if
\begin{equation}
  \label{eq:morse-ineq}
  \frac{\Gamma(1-2k)}{\Gamma(-k-A)}=
\frac{\Gamma(-1-2A-2m)}{\Gamma(-1-2A-m)}>0,
\end{equation}
which holds for even $m$, and fails for odd.

For $\phi_4(x)$ we need $k+A=-m$.  Hence,
\[ \phi_4(x) = \phi\mo(x;-m-A,0,1)=\phi_+(x)\]
and therefore it generates an isospectral deformation.

It follows that the only algebraic deformations of the Morse
potential corresponding to backward transformations, correspond to
the factorization function
\[
  \phi\up{m}\mo(x)=\phi_3(x)\propto e^{(m+A+1)x+\frac{1}{2}e^{-x}}
  L_m^{-2(1+m+A)}(-e^{-x}), \quad A>0,\quad m\mbox{ even.}
\]
The resulting potentials have the form
\begin{equation}
  \label{eq:modmorse}
  U\up{m}\mo(x) = -(A+\tfrac{3}{2}) \e^{-x} +  \tfrac{1}{4}
  \e^{-2x} - 2
  \partial_{xx} \lp \log L_m^{-2(1+m+A)}(-e^{-x})\rp.
\end{equation}
The Darboux transformation corresponds to the operator
\begin{eqnarray*}
  \alpha\mo\up{m} &=& \partial_x -\partial_x(\log \phi\up{m}\mo)   \\
   &=& \partial_x -(1+m+A)+\tfrac{1}{2}e^{-x}
     +\frac{e^{-x}\,r_m'(e^{-x})}{r_m(e^{-x})},
\end{eqnarray*}
where
\[   r_m(z)=L_m^{-2(1+m+A)}(-z).\]
Applying $\alpha\mo\up{m}$ to the bound states
\eref{eq:morse-bound}, we infer that the bound states of
the deformed potential are
\[\psi\up{m}\mop{j}(x)=\lp\frac{e^{(j-A)x-\frac{1}{2}e^{-x}}}{r_m(e^{-x})}\rp
p\up{m}_j(e^{-x}),\]
where
\begin{eqnarray*}
  p\up{m}_j(z)&=&  q_j(z) r_m(z) (z+n-m-2A-1) - z\, q_j'(z)\, r_m(z)
  +
  z\,q_j(z)\, r_m'(z)\\
  q_j(z)&=&  L_j^{2(A-j)}(z).
\end{eqnarray*}
Note that the $p\up{m}_j(z)$ are polynomials of degree $j+m+1$, and
hence, for every $n$, the polynomials
\[1,\;p\up{m}_0(z),\;p\up{m}_1(z),\;\ldots,\;p\up{m}_{n}(z),\]
span a codimension $m$
invariant subspace of $\calP_{m+n+1}(z)$.
We have demonstrated that algebraic deformations of the Morse potential
 are  exactly solvable by polynomials.
\begin{figure}[htbp]
  \begin{center}
    \noindent\psfig{figure=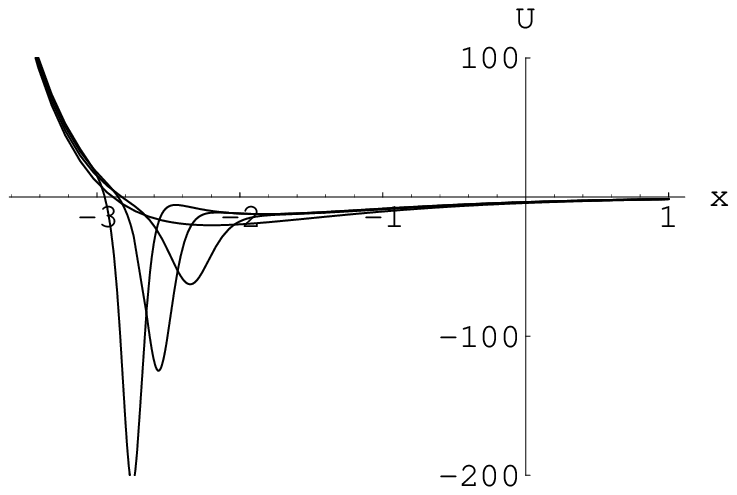,width=4in}
    \caption{Algebraic deformations $U\up{m}(x)$ of the Morse
      potential for $A=2.5$ and $m=0,1,2$ and $3$.}
    \label{fig:morse}
  \end{center}
\end{figure}


\subsection{The hyperbolic P\"oschl-Teller potential.}
The hyperbolic P\"oschl-Teller potential\cite{ptell}, which
includes the class of
reflectionless 1-soliton potentials\cite{matveev},
has the form
\begin{equation}\label{PT}
U\ptl(x)=\tfrac{1}{4}\lp \tfrac{1}{4}- A^2\rp\, \sech^2(\tfrac{x}{2}).
\end{equation}
The general solution\cite[Sec.~2.9]{bateman} of the corresponding
Schr\"odinger equation
\[ -\phi''(x) + U\ptl(x)\phi(x)=-k^2 \phi(x)\]
can be
given as
\begin{eqnarray*}
  \hskip -5em
  \phi\ptl(x;k,C_0,C_1) &=&  \cosh(\tfrac{x}{2})^{\frac{1}{2}-A}\Big\{ C_0\,\,
  F(-\tfrac{A}{2}+\tfrac{1}{4}+k,-\tfrac{A}{2}+\tfrac{1}{4}-
  k,\tfrac{1}{2};-\sinh^2(\tfrac{x}{2}))\\
  && +  C_1\, \sinh(\tfrac{x}{2})\,
  F(-\tfrac{A}{2}+\tfrac{3}{4}+k,-\tfrac{A}{2}+\tfrac{3}{4}-
  k, \tfrac{3}{2};-\sinh^2(\tfrac{x}{2})) \Big\},
\end{eqnarray*}
where $F(a,b,c;z)$ also denotes the analytic continuation of the
hypergeometric function to $\mathrm{Re}(z)<0$.  For $A>1/2$ , the
potential \eref{PT} has $\lceil A-\tfrac12 \rceil$ bound states
\[\psi\ptlp{j}(x),\quad 0\leq j<A-\tfrac12.\]
The even
bound states are
\begin{eqnarray}
  \label{eq:pt-even}
   \psi\ptlp{2i}(x)&\propto&
   \phi\ptl(x;\tfrac{A}{2}-i-\tfrac{1}{4},1,0) \\
   \nonumber
   &\propto &
   \cosh(\tfrac{x}{2})^{\frac{1}{2}-A}\, P_i^{(-\frac{1}{2},-A)}(\cosh x) .
\end{eqnarray}
The odd ones are
\begin{eqnarray}
  \label{eq:pt-odd}
  \hskip -3em
  \psi\ptlp{2i+1}(x) &\propto&
  \phi\ptl(x;\tfrac{A}{2}-i-\tfrac{3}{4},0,1)\\
  \nonumber
  &\propto&
  \sinh(\tfrac{x}{2})\cosh(\tfrac{x}{2})^{\frac{1}{2}-A}\,
  P_i^{(\frac{1}{2},-A)}(\cosh x).
\end{eqnarray}

We  focus on deformations of potentials with bound states
only, i.e., $A>\tfrac{1}{2}$.  The spectral minimum is $-(\tfrac{1}{2}-A)^2$.
For $|k|>\tfrac{1}{2}-A$, the
nodeless solutions of \eref{eq:hoformals}
are given by
\begin{equation}
  \label{eq:pt+-}
  \phi\ptl(x;k,1,t),\quad |t|\leq
    2\,\frac{\Gamma(\frac{3}{4}+k-\frac{A}{2})\Gamma(\frac{3}{4}+k+\frac{A}{2})}{
    \Gamma(\frac{1}{4}+k-\frac{A}{2})
    \Gamma(\frac{1}{4}+k+\frac{A}{2})},
\end{equation}
with the extreme values of $t$  corresponding to $\phi_\pm$
\cite[Sec. 2.3.2, Sec. 2.10]{bateman}.

To obtain algebraic deformations we apply \eref{eq:hgpolys} with
\begin{equation}
  \label{eq:ptparms}
  a = \tfrac{1}{4} + k - \tfrac{A}{2},\quad b=\tfrac{1}{4} -k -
  \tfrac{A}{2},\quad   c = 1-A,
\end{equation}
and consider the four possible factorization
functions
$$\phi_i(x)=\cosh(\tfrac{x}{2})^{\tfrac{1}{2}-A}
f_i(\cosh^2(\tfrac{x}{2})),\quad i=1,2,3,4.$$
We rule out $\phi_2(x)$ and
$\phi_3(x)$ because these are odd functions, and hence have a node.
The factorization functions of the form
\begin{eqnarray*}
  \phi_1(x) &=& \phi\ptl(A,\tfrac{A}{2} - \tfrac{1}{4} -m,1,0)\\
  &\propto&\cosh(\tfrac{x}{2})^{\tfrac{1}{2}-A}
  P_m^{(-\frac12,-A)}(\cosh(x))
\end{eqnarray*}
are nodeless for
$m>A-\tfrac{1}{2}$.
The factorization functions of the form
\begin{eqnarray*}
  \phi_4(x) &=&\phi\ptl(A,-\tfrac{A}{2}-\tfrac{1}{4} - m,1,0)\\
  &\propto& \cosh(\tfrac{x}{2})^{\tfrac{1}{2}+A}P_m^{(-\frac12,A)}(\cosh(x))
\end{eqnarray*}
are nodeless for all $m=0,1,2,\ldots$.

Thus, we see that there are two series of algebraic deformations.
In order to study deformations for all possible $m$, we focus on the
latter series.  The resulting potentials have the form
\begin{equation}
  \label{eq:modptell}
  U\up{m}\ptl(x) =
  -\tfrac{1}{4} (A+\tfrac12)(A+\tfrac32)\, \sech^2(\tfrac{x}{2}) - 2\,
  \partial_{xx} \lp \log
  P_m\up{-\frac{1}{2},A}(\cosh x)\rp.
\end{equation}
\begin{figure}[htbp]
  \begin{center}
    \noindent\psfig{figure=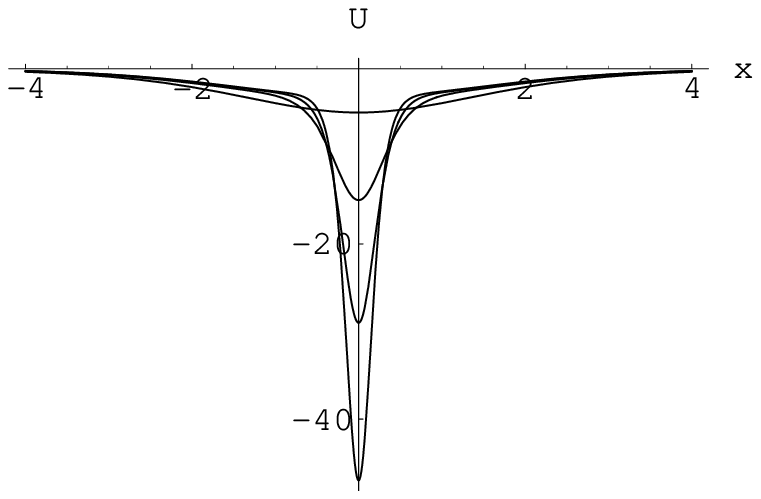,width=4in}
    \caption{ Algebraic deformations $U\up{m}\ptl(x)$  of the hyperbolic
      P\"oschl-Teller potential \eref{eq:modptell}, with $A=4$ and $m=0,1,2$ and $3$.}
    \label{fig:pt}
  \end{center}
\end{figure}
The Darboux transformation corresponds to the operator
\begin{eqnarray*}
  \alpha\ptl\up{m} &=& \partial_x -\partial_x\log \phi_4(x)
\end{eqnarray*}
Applying $\alpha\ptl\up{m}$ to the even bound state functions
\eref{eq:pt-even} yields
\begin{equation}
  \label{eq:pt-defefunc}
    \psi\up{m}\ptlp{2j}(x)=\mu\up{m}(x) \; s_j\up{m}(w)
\end{equation}
where
\begin{eqnarray*}
  \mu\up{m}(x) &=& \frac{\cosh(\frac{x}{2}) \sinh(x)}{(w+1) q_m(w)} \\
  w&=&2z+1 = \cosh(x)\\
  s_j\up{m}(w) &=&   (w+1) \left\{p_j'(w) q_m(w)- q_m'(w)
  p_j(w)\right\} - A\, q_m(w)
  p_j(w), \\
  q_m(w) &=& P_m^{(-\frac12,A)}(w)\\
  p_j(w) &=& P_j^{(-\frac12,-A)}(w).
\end{eqnarray*}
Hence, the deformed operator, conjugated by the gauge factor
$\mu\up{m}(x)$, preserves the codimension $m$ submodule of
$\calP_{m+n}(z)$ spanned by $s_j\up{m}(w),\; j=0,\ldots,n$.  A similar
result holds for the deformation of the odd bound states
\eref{eq:pt-odd}.  Therefore, algebraic deformations of the hyperbolic
P\"oschl-Teller potential are exactly solvable by polynomials.


\section{Exceptional monomial modules.}
In this section we characterize the algebraic structure of the
first-fold deformations $(m=1)$ described in the preceding section.
We will show that this is precisely the class of  exactly
solvable operators that preserves the infinite flag of polynomial
modules
\begin{equation}\label{flag}
\calP\up{1}_0\subset\calP\up{1}_2\subset\calP\up{1}_3\subset\ldots\subset\calP\up{1}_n
\subset\ldots,
\end{equation}
where
\begin{equation}\label{exceptional}
\calP\up{1}_{n} = \lspan\{1,z^2,z^3,\ldots,z^n\},\qquad \calP\up{1}_0 = \lspan\{1\}.
\end{equation}
Will will call such modules {\em exceptional monomial modules}.  They are
exceptional in the sense that the family of second order operators
that leave them invariant is very rich \cite{turbiner2,gkm}. To this
effect, let us begin by the following
\begin{prop}\label{propT}
  A second-order differential operator preserves $\calP\up{1}_n$ if and
  only if it is a linear combination of the following  7 operators:
\begin{eqnarray}
  T\up{+2}_2  &= z^4\partial_{zz} + 2(1-n)z^3\partial_z + n(n-1)z^2\;,\\
  T\up{+1}_2   &= z^3 \partial_{zz} -(n-1) z^2\partial_z ,\\
  T\up{0}_2 &= z^2\partial_{zz} \;,\\
  T\up{-1}_2   &= z\partial_{zz} -\partial_z \;,\\
  T\up{-2}_2   &=\partial_{zz}-2z^{-1}\partial_z\;,\\
  T\up{0}_1 &= z\partial_z\;,\\
  T\up{0}_0 &= 1.
\end{eqnarray}
\end{prop}
A proof of the Propostion is found in \cite{turbiner2,gkm}.
If the linear combination contains the raising operators
$T\up{+2}_2$ and $ T\up{+1}_2$ then the operator will preserve
$P\up{1}_n$ but not the whole flag \eref{flag}. These cases are
called quasi-exactly solvable in the literature
\cite{turbiner1,ko,gko} and will be analyzed in detail in
\cite{gkm}. Since we restrict to exactly solvable cases, we shall
consider only the following linear combination
\begin{equation}\label{T}
T = p_2 T\up{0}_2 + p_1 T\up{-1}_2 + p_0 T\up{-2}_2 + q_2
T\up{0}_1,
\end{equation}
where the additive constant has been neglected.
This can be written as
\begin{equation}\label{Top}
T= P(z) \partial_{zz} + Q(z)
\partial_z,
\end{equation} where
\begin{eqnarray}\label{Pz}
P(z) &= p_2\,z^2+ p_1\, z + p_0,\\
Q(z) &= q_2\,z - p_1 - 2p_0\, z^{-1},
\end{eqnarray}
are quadratic polynomials whose coefficients $p_2, p_1, p_0$ and $
q_2$ are arbitrary real numbers.

The exceptional monomial module $\calP\up{1}_n$ is invariant with
respect to scaling of the $z$ variable. By also allowing rescaling
of the physical variable $x$, it suffices to
consider the following canonical cases.

\begin{table}[htbp]
  \begin{center}
    \begin{tabular}{c|ccc}
      & $P(z)$ & $Q(z)$ & $z(x)$\\
      \hline
      Ia & $ \quad (1-z)(z+2+2A)$  & $q_2 z+2A+1-4(1+A)z^{-1}$  &
      $(\tfrac32+A)\cosh(x)-A-\tfrac12$ \\
      Ib & $ z(1-z)$ & $q_2 z - 1$  & $\cosh^2(\tfrac{x}{2})$\\
      Ic & $ -(1+z^2)$ & $q_2 z +2 z^{-1}$ & $\sinh x$\\
      IIa & $-(z-1)^2$ & $q_2 z -2+2 z^{-1}$ & $-(2A+3)e^{x}+1$ \\
      IIb & $ -z^2$ & $q_2 z$ & $e^{x}$\\
      IIIa & $8(1-z)$ & $q_2 z+8-16 z^{-1}$ & $2x^2+1$\\
      IIIb & $-4z$ & $q_2 z + 1$ & $x^2$ \\
      IV & $-1$ & $q_2 z +2 z^{-1}$ & $x$
    \end{tabular}
    \caption{Second-order operators preserving the exceptional monomial module}
    \label{tab:emm}
  \end{center}
\end{table}
In cases Ib, IIb, IIIb, the operator is of the form shown in
\eref{eq:esform}, and therefore preserves the full $\calP_n$ and not
just the exceptional module $\calP\up{1}_n$.  Thus, these cases
describe undeformed shape-invariant potentials.  Cases Ic and IV
correspond to singular potentials, and will not be discussed further.

 Proposition \ref{prop:zrange} shows that the non-singular potentials
 in case Ia  correspond to
\[ q_2 = \pm(A+\tfrac32)\]
Both possibilities yield the same potential form, so we take the
former. Using \eref{eq:potform} \eref{eq:modptell} we have
\begin{eqnarray*}
 U(x) &=& -\tfrac14(A+\tfrac12)(A+\tfrac32) + 2\,\frac{k \cosh
  x-1}{(\cosh x-k)^2},\quad k=\frac{2A+1}{2A+3},\\
  &=& U\up{1}\ptl(x)+(\tfrac54+\tfrac{A}{2})^2.
\end{eqnarray*}

For similar reasons, for case IIIa potentials we must take
$q_2 = 4$.  Using \eref{eq:potform} \eref{eq:modhosc} we obtain
\begin{eqnarray*}
  U(x) &=& 3+x^2 +\frac{8}{2x^2+1}-\frac{16}{(2x^2+1)^2},\\
  &=& U\up{1}\ho(x)+5
\end{eqnarray*}

For the case IIa, we translate the $x$ variable and do a spectral
shift to set $q_2=2A+3$.  In this way, \eref{eq:potform}
\eref{eq:modmorse} yields
\begin{eqnarray*}
  U(x) &=& (2+A)^2 +\tfrac14 e^{-2x}- (A+\tfrac32) e^{-x} +
  \frac{2ke^{x}}{(1-ke^x)^2},\quad k = 2A+3,\\
  &=&U\up{1}\mo(x)+(2+A)^2.
\end{eqnarray*}
We should
note that the above potential form is non-singular  only if
$2A+3<0$.  This is unavoidable, in as much as we showed in section
\ref{sub:morse} that the odd deformation of Morse potentials  with
bound states produce
singular potentials.

In summary, we have demonstrated that non-singular Hamiltonians that
are exactly solvable by an infinite flag of exceptional monomial
modules and not by the ordinary flag \eref{pflag} are precisely the
first-fold algebraic deformations of the non-singular shape-invariant
potentials.

Although these new potentials preserve a full flag of polynomial
subspaces, and therefore are exactly solvable in the sense
defined by Turbiner in \cite{turbiner1}, they do not possess a
hidden $\mathfrak{sl}(2)$ symmetry algebra structure. This shows
that the exactly solvable class is wider than the Lie-algebraic
one.

\ack The research of DGU is supported in part by a CRM-ISM
Postdoctoral Fellowship and the Spanish Ministry of Education
under grant EX2002-0176. The research of NK and RM is supported by
the National Science and Engineering Research Council of Canada.
The authors would like to thank Prof. Gonz\'alez-L\'opez and Prof.
Gesztesy for interesting discussions, as well as the referees,
who made very interesting remarks on the first version of the
paper.


\vskip 1cm


\begin{thebibliography}{99}

\bibitem{darboux} Darboux G, \emph{Th\'eorie G\'en\'erale des
    Surfaces}, vol. II, Gauthier-Villars, 1888.
\bibitem{jacobi} Jacobi CG, 1837 \emph{J. Reine Angew. Math.} {\bf 17}, 68.
\bibitem{schrodinger}
Schr\"odinger E 1941 {\it Proc. Roy. Irish Acad.} {\bf 47} A, 53
({\it Preprint} physics/9910003).
\bibitem{infeld-hull}
Infeld L and Hull T E 1951 {\it Rev. Mod. Phys.} {\bf 23} 21.
\bibitem{cooper}
Cooper F, Khare A and Sukhatme U 1995 {\it Phys. Rep.} {\bf 251}
267.
\bibitem{deift} Deift P and Trubowitz E 1979 {\it Duke Math J.} {\bf
    45}, 267.
\bibitem{gesztesy} Gesztesy F, Simon B and Teschl G 1996 {\it
    J. d'Analyse Math.} {\bf 70}, 267
\bibitem{calogero} Calogero F and Degasperis A  1982 {\it  Spectral
    transform and solitons I}, Studies in Mathematics and its
    Applications (New York:Elsevier).

\bibitem{sukumar} Sukumar CV 1985 {\it J. Phys. A} {\bf 18} 2917.
\bibitem{sparenberg}Sparenberg J-M and Baye D 1995 {\it J. Phys. A}
    {\bf 28} 5079.
\bibitem{bagrov1}
Bagrov V G and Samsonov B F 1995 {\it Theoret. and Math. Phys.}
{\bf  104} 1051.
\bibitem{gendenshtein} Gendenshtein L 1983 {\it JETP Lett} {\bf 38}  356.
\bibitem{mielnik}  Mielnik B 1984 {\it J. Math. Phys.}  {\bf 25}
  3387.
\bibitem{levai} L\'evai G, Baye D and  Sparenberg J-M 1997 {\it
    J. Phys. A}  {\bf 30} 8257
\bibitem{turbiner1}
Turbiner A V 1988 {\it Commun. Math. Phys.} {\bf 118} 467.
\bibitem{ko}
Kamran N and Olver P J 1990 {\it J. Math. Anal. Appl.} {\bf 145}
342.
\bibitem{gko}
Gonz\'alez-Lopez A,  Kamran N and  Olver P J 1993 {\it Commun.
Math. Phys.} {\bf 153} 117.
\bibitem{morse}
Morse P M 1929 {\it Phys. Rev.} {\bf 57} 57.
\bibitem{ptell}
P\"oschl G and Teller E 1933 {\it Z. Physik} {\bf 83} 143.
\bibitem{milson} Milson R 1998 {\it Internat. J. Theoret. Phys.}  {\bf 37} 1735.
\bibitem{ggr} G\'omez-Ullate D, Gonz\'alez-L\'opez A and
Rodr\'{\i}guez M A 2000 {\em J. Phys. A} {\bf 33} 7305.
\bibitem{turbiner2}
Post G and Turbiner A V 1995 {\it Russian J. Math. Phys.} {\bf 3}
113.
\bibitem{finkelkamran}
Finkel F and Kamran N 1998 {\it Adv. in Applied Math.} {\bf 20}
300.
\bibitem{baye} Baye D, Sparenberg J-M and L\'evai G 1997 {\em Inverse
    and Algebraic Quantum Scattering Theory (Lecture notes in Physics
    488)} ed B Apagyi, G Endr\'edi and P L\'evay (Berlin: Springer) p
    295
\bibitem{gkm} G\'omez-Ullate D,  Kamran N  and Milson R, in preparation.
\bibitem{gt} Gonz\'alez-L\'opez A  and Tanaka T,
  hep-th/0307094.
\bibitem{shifman} Shifman  M 1989 {\em Int. J. Modern Phys. A} {\bf 4} 3311.
\bibitem{schminke}Schminke U W  1978, {\it Proc. Roy. Soc. Edinburgh
     Sec. A} {\bf 80}, 67.

\bibitem{bateman}
Erd\'elyi A et al. 1953 {\em Higher Transcendental Functions, Vol. I},
(New York:McGraw-Hill).
\bibitem{dubov} Dubov S Y,  Eleonskii V M  and Kulagin N E 1992,
  {\it Sov. Phys. JETP} {\bf 75} 446.
\bibitem{bagrov2} Bagrov V G and Samsonov B F 1997 {\it Pramana J. Phys.} {\bf
    49} 563.

\bibitem{matveev}  Matveev V and Salle M A 1991 {\it Darboux transformations and
solitons}, Springer Series in Nonlinear Dynamics (Berlin:Springer)

\end{thebibliography}
\end{document}